\documentclass[runningheads]{llncs}
\usepackage{graphicx}
\usepackage{hyperref}
\usepackage{comment}
\usepackage{breakcites}
\begin{document}
\title{From Keywords to Structured Summaries: Streamlining Scholarly Information Access}
\titlerunning{Streamlining Scholarly Knowledge Access}

%
\author{Mahsa Shamsabadi \and
Jennifer D'Souza\orcidID{0000-0002-6616-9509}}
\authorrunning{Shamsabadi and D'Souza}
%
\institute{TIB Leibniz Information Centre for Science and Technology, Hannover, Germany
\email{jennifer.dsouza@tib.eu}}
\maketitle              
\begin{abstract}

This paper highlights the growing importance of information retrieval (IR) engines in the scientific community, addressing the inefficiency of traditional keyword-based search engines due to the rising volume of publications. The proposed solution involves structured records, underpinning advanced information technology (IT) tools, including visualization dashboards, to revolutionize how researchers access and filter articles, replacing the traditional text-heavy approach. This vision is exemplified through a proof of concept centered on the ``reproductive number estimate of infectious diseases'' research theme, using a fine-tuned large language model (LLM) to automate the creation of structured records to populate a backend database that now goes beyond keywords. The result is a next-generation information access system as an IR method accessible at \url{https://orkg.org/usecases/r0-estimates}.

\end{abstract}

\section{Introduction}






The proliferation of scientific literature demands a reevaluation of scholarly article information retrieval (IR) engines~\cite{fortunato2018science,bornmann2021growth}. The rapidly expanding volume of scientific publications across various domains~\cite{altbach2019too,horbach2020pandemic} has led to an urgent need for more efficient and precise methods of accessing and filtering these vast knowledge stores. Traditional keyword-based search engines, which once served as the primary means of IR, are proving inadequate to help researchers keep track of the fast-paced scientific progress. As a result, there is a growing demand for innovative approaches, such as structured scholarly content representations~\cite{ermakova2018abstract,fontelo2013comparing} and advanced machine learning algorithms~\cite{ammar2018construction,pride2023core}, going beyond keywords, facilitating the application of smarter information technology (IT) and access tools to assist the researcher to effectively navigate the text-heavy landscape of scholarly literature. Projects like the \href{https://orkg.org/}{Open Research Knowledge Graph} (ORKG)~\cite{auer2020improving} are driving the shift towards fine-grained structured knowledge representations, enabling smarter views~\cite{smartreviews} and comparisons of research aspects~\cite{oelen2019comparing}. This paradigm shift not only promises to streamline the research process but also holds the potential to democratize access to scientific knowledge, making it more accessible to a broader audience, including non-experts seeking insights into cutting-edge research.

Traditional scholarly article IR engines rely on keywords, bibliometrics, and citation graphs, resulting in ranked lists of relevant publications. However, this approach necessitates time-intensive cognitive tie-ups for the researcher to navigate and filter the text-heavy results. This paper's central research agenda is to improve access and filtering of scholarly knowledge based on uniformly structured scholarly content. The rise of structured scholarly knowledge, seen in projects like the Semantic Scholar Literature Graph~\cite{ammar2018construction} and Open Research Knowledge Graph (ORKG)~\cite{auer2020improving}, signals the need to rethink scholarly information systems with next-generation IT. Our aim is to use IT to streamline scholarly content access, reducing cognitive load on researchers dealing with text-heavy models of filtering publications in traditional IR platforms. To achieve this, we explore the concept of dashboards, which provide easily understandable views of crucial information via visual representations, aiding decision-making~\cite{santos2017data}. A knowledge dashboard on IR platforms, based on structured scholarly knowledge facets, would simplify research by replacing text-heavy ranked lists with smart IT-generated charted summaries for filtering publications and research discovery. The device of dashboards is not new in representing scientific information~\cite{orkg-dashboard,smartcity-dashboard}. Khodaveisi et al.~\cite{covid19-scoping-review} examined 26 recently introduced dashboards during the Covid-19 pandemic, focusing on their technical features~\cite{salehi2021synergetic,yang2021using,zhu2021dashboard,aristizabal2023interactive,hodgson2020covid,ravinder2020adaptive,ulahannan2020citizen,wissel2020interactive,peddireddy20205vs,bae2020information,florez2020online,pathak2020racial,ibrahim2020rapid,chande2020real,marivate2020use,hohl2020daily,carroll2021using,marques2021dashboard,hyman2021data,clement2020interactive,dixon2021leveraging,arias2021outbreak,chauhan2021understanding}. These dashboards mainly served citizen science by tracking COVID-19 cases, analyzing epidemiological trends, and aiding decision-making, drawing data from public databases like the World Health Organization (WHO) and Johns Hopkins, or electronic health records at the hospital organization level. In contrast, we aim to apply novel IT to structured facets of scholarly knowledge itself, streamlining scholarly knowledge access as next-generation information systems based on data extraction mechanisms supported by LLMs.


In pursuit of our vision, this short paper demonstrates a proof-of-concept (POC) over a collection of 2000 articles on the ``reproductive number estimate of infectious diseases'' research theme. The reproductive number estimate for infectious diseases, often denoted as ``$R0$,'' is a measurement indicating the average number of secondary cases generated by a single infected individual in a population~\cite{gordis2013epidemiology}. We adopt the \textsc{orkg-R0} semantic model~\cite{orkg-ro}, defining six crucial properties to structure the salient content for the research theme: the \textit{disease name} under investigation, the study \textit{location}, the study \textit{date}, the \textit{$R0$ value}, the \textit{\%Confidence Interval values}, and the computation \textit{method} for $R0$. In scholarly knowledge organization, properties are research-theme-specific.However the forthcoming discussion on smart IT dashboards applied over structured scholarly content applies generally to various themes and properties. More than that, well-defined properties enable the comparison among different research works so one can determine which works to read further w.r.t. the compared property value. In our POC, we treat the properties as indicators of interest to researchers when filtering for scholarly information on the theme. W.r.t. next-gen scholarly information systems we posit that they should address central research questions that enable smart filtering of relevant publications. Thus we define four research questions (\textbf{RQ}s) for them. \textbf{RQ1}: what are the max $R0$ estimates reported for the diseases in our database? \textbf{RQ2}: For a chosen disease, how many studies have been reported across study locations? \textbf{RQ3}: what is the min and max $R0$ for a disease across study locations? \textbf{RQ4}: locate on the world map where the diseases have been studied? In our POC, these RQs are represented as chart summaries in a dashboard, enhancing publication and research filtering. This IT-driven approach deconstructs scholarly knowledge into user-friendly visuals, improving the filtering process beyond keyword-based methods and empowering researchers. The rest of the paper explains the technical details of the various modules that bring our POC to life.

\section{The Automatic Knowledge Structuring Model}

We employ the ORKG-FLAN-T5 R0 large language model (LLM) ~\cite{shamsabadi-etal-2024-large} from Hugging Face, available at \href{https://huggingface.co/orkg/R0_contribution_IE}{link}. This LLM, based on the T5 architecture~\cite{t5}, is finely tuned for precise extraction of epidemiological data, particularly for generating structured research contribution summaries related to the $R0$ estimate in infectious diseases. It aids in systematically extracting structured epidemiological information from scientific literature, benefiting the fields of epidemiology and virology by addressing the specific challenge of $R0$ estimation. The question at hand is: what are the key properties for structuring a summary of $R0$ estimate research in infectious diseases? To address this, an expert semantic modeler created a \href{https://orkg.org/comparison/R44930/}{research comparison} based on property-value pairs for Covid-19 $R0$ estimate contributions from 30 abstracts, resulting in six properties: \textit{disease name}, \textit{location}, \textit{date}, \textit{R0 value}, \textit{\%Confidence Interval (CI) values}, and \textit{method}. This structuring, known as \textsc{orkg-R0}, aims for a balanced approach between generality and specificity in organizing $R0$ estimate research. Subsequently, a \href{https://zenodo.org/records/8068442}{larger corpus} of 1,500 articles underwent annotation for machine learning training. Each article's title and abstract were paired with one or more structured summaries depending on the number of $R0$ values reported.

The ORKG-FLAN-T5 R0 model~\cite{shamsabadi-etal-2024-large} is an instruction fine-tuned variant of \href{https://github.com/google-research/t5x/blob/main/docs/models.md#flan-t5-checkpoints}{FLAN-T5 Large} (780 M) using instruction finetuning as an incremental progression of the instruction-tuning paradigm introduced as FLAN (Finetuned Language Net)~\cite{t5,flan,flan-t5,flan-collection}. The model is trained to respond to the structuring task objective formulated as the question: \textit{What are the values for the following properties of the basic reproduction number estimate (R0): disease name, location, date, R0 value, \%CI values, and method?} The model processes a paper's title and abstract, producing a structured description. A sample of the LLM versus gold-standard human responses can be found online at this \href{https://scinext-project.github.io/#/r0-estimates}{URL}.

\section{Our Scholarly Articles Collection}

Our POC article collection focused specifically on relevant studies related to our chosen research theme of $R0$ estimates. We obtained these articles through keyword-based searches in the  \href{https://pubmed.ncbi.nlm.nih.gov/}{PubMed} database, the largest source of biomedical and life sciences research articles. Our latest search was conducted on September 13, 2023, to compile the POC collection. The exact search query applied was the following: \texttt{(basic reproduction number[TIAB] OR basic reproductive number[TIAB] OR basic reproduction ratio[TIAB] OR basic reproductive rate[TIAB] OR basic reproductive ratio[TIAB] OR basic reproduction rate[TIAB] OR  R0[TIAB])  NOT (R0 resection OR cancer)}. The search query aimed to retrieve papers containing any synonym variant of $R0$ estimate in the title or abstract, yielding 7,127 articles. The search results were exported as a CSV file, including metadata fields like PubMed ID, publication date, title, and abstract.

\subsection{Our Structured Scholarly Articles Collection}
\label{subsec:collection}

The initial keyword-based article collection did not constitute the final dataset for our IR platform, as articles not reporting an R0 value needed to be filtered out. To refine our collection, we employed the ORKG-FLAN-T5 R0 LLM, queried through Python with the Transformers library. This model, trained to provide structured JSON descriptions for articles reporting R0 estimates, excluded unanswerable cases. Subsequently, unanswerable records were removed, resulting in 2,051 remaining articles with 2,736 structured descriptions. These JSON response objects were reorganized into columns based on six properties, and \textit{location} property values were normalized using \href{http://www.geonames.org/export/web-services.html}{GeoNames} and other sources, extrapolated into a new column. For $R0$ values reported in ranges, preprocessing split them into minimum and maximum values. The processed data was imported to a PostgreSQL 16 database, serving as the backend storage. A query for the top 20 most represented infectious diseases in our database is shown in \autoref{tab:diseases}. Notably, the high precision demonstrated by the LLM ensured that all the top reported diseases are indeed ascertained infectious diseases. Geographically, our database includes studies from all seven continents, with the top 20 countries being: China, India, United States, Italy, Brazil, Japan, South Korea, Iran, United Kingdom, Saudi Arabia, Spain, Pakistan, Germany, Bangladesh, Canada, Nigeria, France, Colombia, Netherlands, and Taiwan.

\begin{table}[!htb]
\centering
 \caption{The top 20 infectious disease names (and number of papers) in our dataset.} \label{tab:diseases}
 \begin{tabular}{|l | l | l | l |} 
 \hline
 covid-19 (1002) & mers-cov (21) & measles (15) & hepatitis c (8) \\ 
 dengue (41) & cholera (18) & hepatitis b (12) & tuberculosis (8) \\
 influenza (29) & zika (18) & zika virus (12) & monkeypox (8) \\
 hiv (23) & african swine fever (17) & ebola (11) & west nile virus (7) \\
 sars (22) & ebola (17) & hand, foot, and mouth disease (8) & malaria (7) \\
 \hline
 \end{tabular}
\end{table}



\section{A Next-Generation Scholarly IR Platform}

\begin{figure}[!tb]
\includegraphics[width=\textwidth]{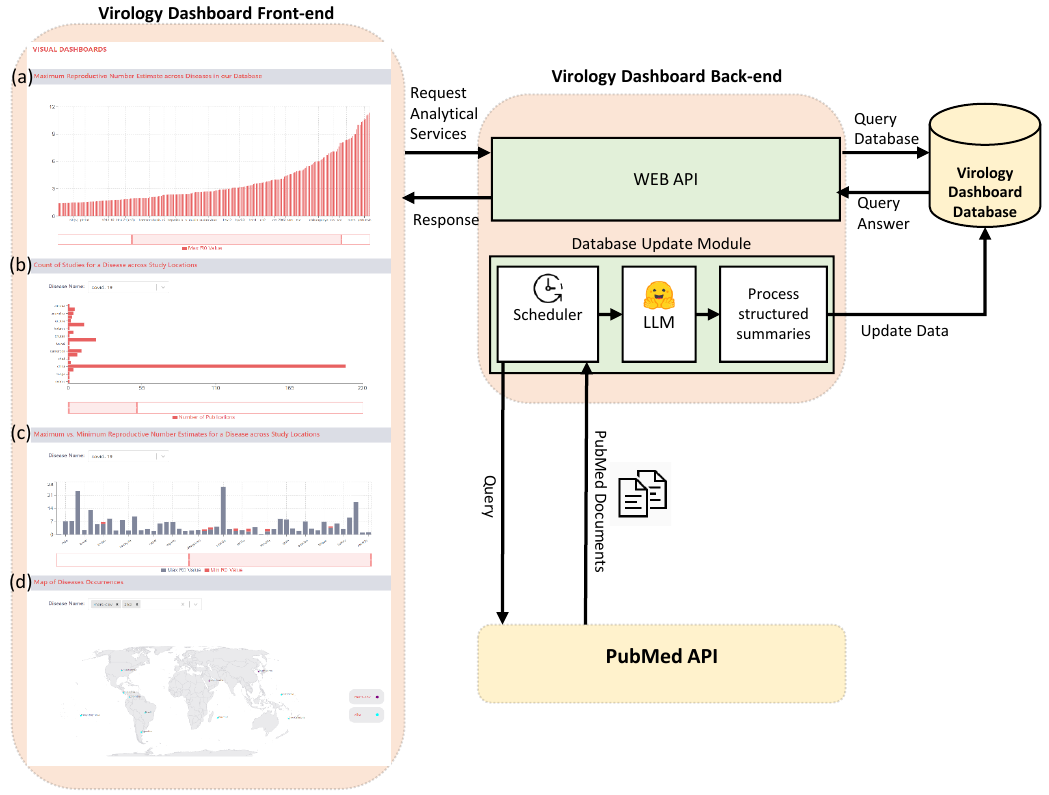}
\caption{\textbf{(Left image)} A visual analytical dashboard in our \href{https://orkg.org/usecases/r0-estimates}{next-gen IR platform} provides charts (a), (b), (c), (d) to help researchers make informed article filtering decisions. \textbf{(Right image)} The backend workflow, managed by a web API, handles database interactions for frontend rendering. It incorporates a scheduler for database updates, with LLM queries supplying structured scholarly knowledge before each update.}\label{fig:platform}
\vspace{-20pt}
\end{figure}

We introduce a next-generation scholarly IR platform dashboard as a proof of concept (POC) for our selected theme on ``reproductive number estimates of infectious diseases.'' The aim is to streamline scholarly knowledge access with advanced frontend IT. This is addressed by breaking down complex scholarly information into four user-friendly visual summaries that consolidate properties from structured knowledge. Each visual provides a concise summary, addressing a specific RQ as an assistant to the researcher, thus enabling to make more informed decisions when filtering for papers. The platform is freely accessible as a web-based application at the following URL: \url{https://orkg.org/usecases/r0-estimates}. Furthermore, just its visualization dashboard widget and underlying workflow are displayed in \autoref{fig:platform}. In this technical workflow, the \href{https://anonymous.4open.science/r/virology-dashboard-frontend-715B/README.md}{frontend} communicates with the \href{https://anonymous.4open.science/r/virology-dashboard-backend-73E1/README.md}{backend} through a Web API for database queries and data retrieval. Python scripts manage service requests, interpreting component queries and directing database queries. A Python script \href{https://anonymous.4open.science/r/virology-dashboard-backend-73E1/virology_contributions_api/scheduler/scheduler.py}{scheduler} periodically updates the database with new articles, querying PubMed for the initial article collection and following the processing cycle as outlined in \autoref{subsec:collection} before updating the database with structured summaries. Our workflow maximizes the use of cutting-edge technology, including an optimized next-generation LLM.





\begin{figure}[!htb]
\includegraphics[width=\textwidth]{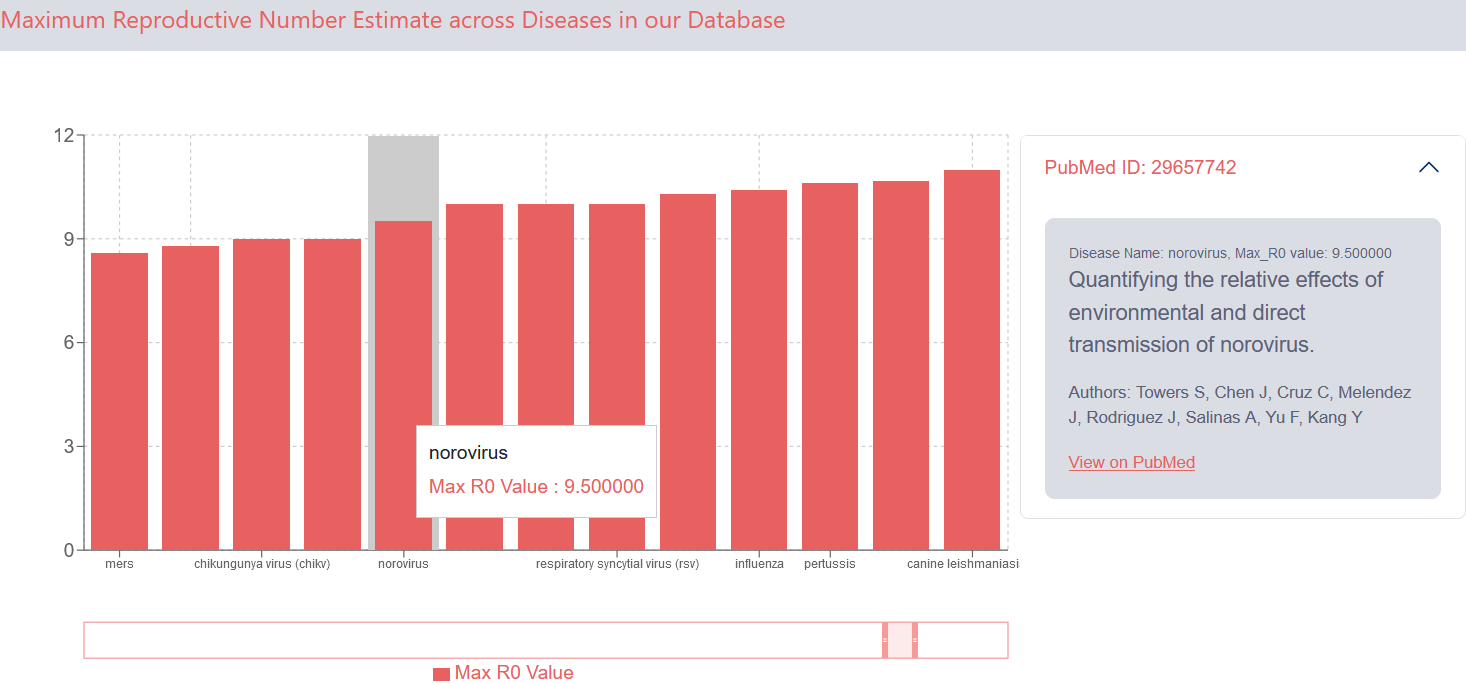}
\caption{A closer look at chart (a) in \autoref{fig:platform}. This chart was designed to address the research question: ``What are the maximum R0 estimates reported for the diseases?'' to support advanced scholarly publication filtering. The y-axis displays max R0 values, while the x-axis shows various infectious diseases in our database. The chart facilitates filtering by allowing selection of the R0 value range to display. Additionally, clicking on each bar reveals the list of the publications whose data underlies the bar, with clickable links redirecting to the respective articles on PubMed.} \label{fig:rq1}
\end{figure}

\begin{figure}[!htb]
\includegraphics[width=\textwidth]{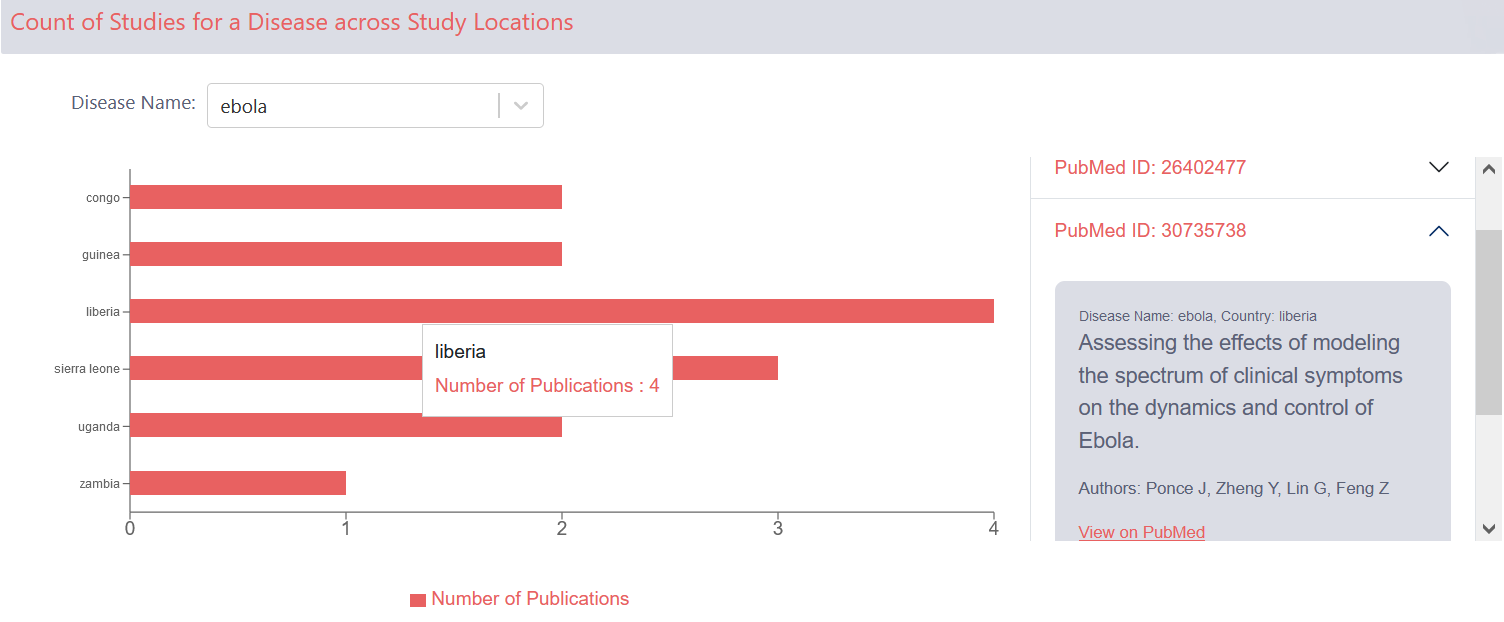}
\caption{A closer look at chart (b) in \autoref{fig:platform}. This chart was designed to address the research question: ``For a chosen disease, how many studies have been reported across study locations?'' The y-axis represents various study locations and the x-axis denotes the number of studies found for that disease in our database. The figure presents the results for disease name Ebola and shows that it was studied in Congo, Guinea, Liberia, Sierra Leono, Uganda, and Zambia. Clicking on each bar reveals the list of the publications whose data underlies the bar, with clickable links redirecting to the respective articles on PubMed.} \label{fig:rq2}
\end{figure}


\subsection{Charting the data: collating, summarizing, and reporting}

Our IR platform comprises three main components: 1) a statistics snapshot displaying total papers, structured knowledge, infectious diseases, and locations in our database, and 2) typical to IR engines, a standard paper listing in a table with keyword-based fields, from which a selected subset is filtered or returned. This is built using the JavaScript data tables library, \href{https://github.com/ag-grid/ag-grid}{ag-grid}. And, 3) a visual analytical dashboard widget featuring four charts that address the previously introduced four research questions. This process involves \textit{collating} relevant structured data properties, selecting the most suitable chart from the React \href{https://recharts.org/en-US/examples}{chart library} that would best \textit{summarize} the response, and creating a query to retrieve the necessary data from the structured database to \textit{report} the visual answer. Each RQ is accompanied by a visual chart. \textbf{RQ1} asks, ``What are the maximum $R0$ estimates reported for diseases in our database?'' A ``\href{https://recharts.org/en-US/examples/SimpleBarChart}{bar chart}'' displays unique diseases on the x-axis and their max $R0$ values on the y-axis, where hovering over a bar shows the disease and its max $R0$ value. This chart, which can be adjusted to focus on specific $R0$ ranges, simplifies the otherwise daunting task of manually comparing $R0$ estimates across thousands of papers. \autoref{fig:rq1} showcases this chart and introduces a feature where clicking on a bar reveals a relevant list of publications that contributed to the bar's data, leading researchers directly to the papers on PubMed. In this way, our approach enhances scholarly information access in ways traditional information retrieval engines do not. \textbf{RQ2} explores, ``For a chosen disease, how many studies have been reported across study locations?'' It is visualized through a ``\href{https://recharts.org/en-US/examples/VerticalComposedChart}{vertical composed chart},'' with study locations on the y-axis and study counts on the x-axis, filtering data by the entered disease name. This extends RQ1 by allowing researchers to investigate the geographic distribution and volume of studies on a particular disease. \autoref{fig:rq2} zooms in on this chart for the chosen disease Ebola, where selecting the Liberia location studies bar shows relevant publications, redirecting users to PubMed papers. \textbf{RQ3}: what is the min and max $R0$ for a disease across study locations? Implemented as a ``\href{https://recharts.org/en-US/examples/StackedBarChart}{stacked bar chart}'' to filter disease-specific min and max $R0$ values across various locations. \textbf{RQ4}: locate on the world map where the diseases have been studied? Enables comparison of up to three disease studies on a world map through a ``\href{https://github.com/zcreativelabs/react-simple-maps}{composable map chart}.'' Each of these charts make possible to alleviate the tedium of traditional IR in filtering for relevant scholarly articles.

Each visual offers interactive disease selection filters and serves as an informed scholarly article filtering option. When enabled, it displays filtered studies accessible directly on PubMed. Our next-gen IR platform simplifies scholarly knowledge exploration, potentially benefiting citizen science.

\subsection{User study}

As a last step, since the platform was developed mainly by computer scientists, we aimed to assess the platform's utility to virologists via a survey. The survey (\url{https://forms.gle/3KAoty4YgUUWsUNt8}) assessed the effectiveness of the dashboard. Four participants evaluated three specific dashboards and responded to two general questions. Using a rating scale from 1 (Not Helpful) to 5 (Very Helpful), participants provided feedback on the dashboards. For each dashboard related question, optional feedback was also solicited. For RQ1, 2 responses chose rating 3 and 2 responses chose rating 5. Here a feedback comment we got was: ``Alternatives with Mean and Median would provide a lot of insight into whether outliers are true or model-driven logical errors.'' RQ2 saw ratings of 4 (2 responses) and 5 (2 responses), with a call for expanding publication database sources for more comprehensive results. RQ3, i.e. showing min $R0$ and max $R0$ for a chosen disease across study locations, received more varied ratings from 2 to 5. Specifically, one respondent rated it as 2, two respondents rated it as 4, and one respondent rated it as 5, indicating a range of opinions on its effectiveness, with concerns raised about outlier display in textual feedback. Regarding the overall platform's assistance in filtering scholarly knowledge compared to keyword-based methods (rated from 1=Extremely Disagree to 5=Extremely Agree), strong agreement (1 respondent rated 4 and three respondents rated 5) indicated significant improvement over traditional methods. In terms of recommending the IR solution to a colleague (rated from 1=Not Likely to 5=Very Likely), high ratings (three respondents rated 4 and 1 respondent rated 5) suggest strong endorsement within virology research.

Our platform sets a benchmark for future advancements, guiding computer scientists keen on improving digital library sciences towards creating innovative information access solutions. These are powered by structured scholarly content representations, enabled by LLMs, and offer advanced knowledge filtering options, implemented via next-gen IT tailored to specific research questions across scientific domains.

\section{Conclusion}

In this paper, we introduced a proof-of-concept for an advanced scholarly IR engine that enhances information access and reduces the cognitive load posed by text-heavy, keyword-based search results in traditional IR. In traditional IR, researchers often need to make mental notes to filter papers for reading, akin to searching for a needle in a haystack due to exponential publication rates. This underscores the urgency to rethink traditional keyword-based scholarly IR by explicitly modeling salient research aspects, making them machine-actionable. Our approach paves the way for next-generation IT filtering visual assistants, streamlining information access in scholarly research.

\bibliographystyle{splncs04}
\bibliography{mybib}

\end{document}